\documentclass[prb,twocolumn,showpacs,floatfix,superscriptaddress,amsmath,amssymb]{revtex4}
\usepackage{graphicx}
\usepackage{graphics}
\usepackage{color}
\usepackage{pst-all}

\newcommand{\beq}{\begin{equation}}
\newcommand{\eeq}{\end{equation}}
\newcommand{\beqa}{\begin{eqnarray}}
\newcommand{\eeqa}{\end{eqnarray}}
\newcommand{\nn}{\nonumber}
\newcommand{\spin}{\vec{S}}

\begin{document}

\title{Quantum phase transitions in trimerized zig-zag spin ladders}


\author{ H.D.\ Rosales}
\affiliation{Departamento de F\'{\i}sica, Universidad Nacional de La Plata, C.C.\ 67, 1900 La
Plata, Argentina.}

\author{ D.C.\ Cabra }
\affiliation{Departamento de F\'{\i}sica, Universidad Nacional de La Plata, C.C.\ 67, 1900 La
Plata, Argentina.} \affiliation{Laboratoire de Physique Th\'{e}orique, Universit\'{e} Louis
Pasteur, 3 Rue de l'Universit\'{e}, 67084 Strasbourg, C\'edex, France.} \affiliation{Facultad de
Ingenier\'\i a, Universidad Nacional de Lomas de Zamora, Cno.\ de Cintura y Juan XXIII, (1832)
Lomas de Zamora, Argentina.}

\author{ M.D.\ Grynberg}
\affiliation{Departamento de F\'{\i}sica, Universidad Nacional de La Plata, C.C.\ 67, 1900 La
Plata, Argentina.}

\author{ G.L.\ Rossini}
\affiliation{Departamento de F\'{\i}sica, Universidad Nacional de La Plata, C.C.\ 67, 1900 La
Plata, Argentina.}

\author{ T.\ Vekua}
\affiliation{Laboratoire de Physique Th\'{e}orique, Universit\'{e} Louis Pasteur, 3 Rue de
l'Universit\'{e}, 67084 Strasbourg, C\'edex, France.}
\affiliation{Laboratoire de Physique Th\'eorique et Mod\`eles
  Statistiques, Universit\'e Paris Sud, 91405 Orsay Cedex, France}

\date{March 6, 2007}

\begin{abstract}
We analyze the effects of a trimerized modulation in a quantum spin $S=\frac12$ zig-zag ladder at the
magnetization plateau $M=1/3$. Such periodicity is argued to be stemmed from lattice
deformations by phonons. The interplay
between frustration and exchange modulation is well described by an effective
triple sine-Gordon field theory close to the homogeneous ladder and by block-spin perturbation theory
in the weakly coupled trimers regime.
The characteristic triple degeneracy of the ground state for homogeneous ladders gives
place to modulation driven quantum phase transitions, leading to a rich phase diagram
including up-up-down, quantum plateau and gapless plateau states.
\end{abstract}

\maketitle

\section{Introduction}

Frustrated quasi one dimensional antiferromagnetic spin $S=\frac12$ systems have been extensively
studied in the last years. One of the most discussed and paradigmatic models is the $J_1-J_2$
zig-zag ladder, which apart from the theoretical interest caused by the frustration is believed to
be the relevant starting point for describing magnetic excitations of a number of real quasi one
dimensional materials \cite{Hase+04}.
Among others, attention is focused on
$\mathrm{CuGeO_3}$ where, along with frustration, spontaneous spin-Peierls exchange modulation caused by the
coupling between spins and phonons plays a crucial role \cite{Hase1993}.
Due to the interest in the above
material the interplay between exchange modulation and frustration has stimulated considerable
efforts in purely spin systems as well, with studies concentrating on the zero magnetization
case \cite{Affleck97,Affleck98,Bouzerar,Dobry,Knetter99}.

Recently it was shown that phonons in frustrated zig-zag ladders can open plateaus not only at zero
magnetization but at other rational values $M$ of saturation (M=1/3, 1/2...) \cite{Vekua06}.
Spontaneous exchange modulations in such situations could also take place, with a spatial pattern associated to
the (generally broken) ground state translational symmetry.

\begin{figure}[htbp]
    \centering
        \includegraphics[width=0.35\textwidth]{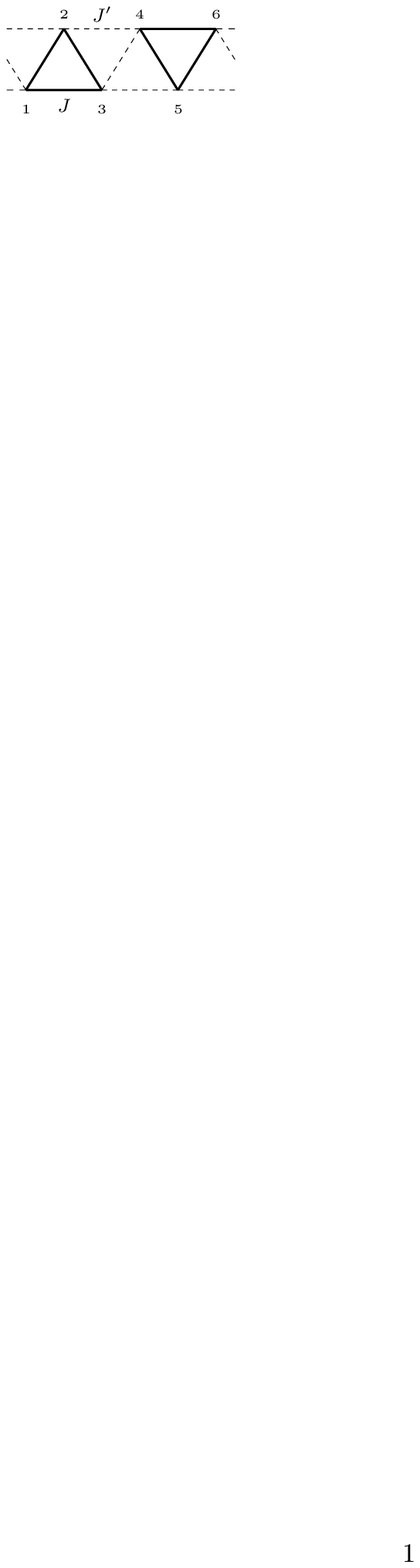}
    \caption{Schematic description of the trimerized zig-zag spin ladder. Both nearest and next nearest
    couplings are modulated.}
    \label{cadena}
\end{figure}

In this work we will concentrate on
the interplay between exchange modulation and frustration in a pure zig-zag spin ladder
at the M=1/3 plateau state.
This system, in absence of modulation, is known to exhibit a magnetization plateau
when the next nearest neighbour (NNN) coupling
$J_2$ is large enough with respect to the nearest neighbour (NN) coupling $J_1$
\cite{Okunishi2003,Lecheminant2004}.
Moreover, the ground state is three-fold degenerate and translation symmetry by one lattice spacing
is spontaneously broken to an up-up-down configuration \cite{Okunishi2003}.
A natural elastic deformation is then given by a period three pattern
\beq
u_i = \delta \sin(\frac{2\pi}{3} i - \Phi_0 ),
\label{deformation}
\eeq
where $u_i$ is a relevant scalar coordinate describing the displacement of the $i$th ion
sequentially numbered on a one dimensional chain.
We will pay attention to the most symmetric situation $J_1=J_2$, and a deformation
with amplitude $\delta>0$ and phase $\Phi_0=- \frac{\pi}{3}$, that causes
site $1$ and $3$ to move closer to site $2$ and so on.
We assume that magnetic exchange couplings vary linearly with the relative displacement to estimate
the modulation on both NN and NNN couplings. In this way,
sites $1,2,3$ are magnetically coupled by equally enhanced exchanges $J=J_1+\lambda$
forming {\em equilateral triangular trimers} as shown in Fig.\ \ref{cadena}.
Couplings drawn with dashed lines are weaker
than $J_1$,
which for simplicity we represent as $J'=J_1-\lambda$.

Notice that this deformation selects triangles with stronger
couplings as basic frustrated units, with weaker couplings between them; in a limiting situation,
that we will also consider, the system is formed by weakly coupled triangular trimers where strong frustration
enhances plateau formation and directly relates to high degeneracy of the ground state.
On the other hand, a deformation with $\delta < 0$
(stronger couplings along dashed lines in Fig.\ \ref{cadena}) modifies the system towards a single homogeneous
spin chain with modulated weaker couplings between non NN, up to fourth neighbours;
the corresponding limiting situation ($J/J'=0$) is just a usual spin chain running along
the dashed line, where no plateau would be
observed in the magnetization curve.

Guided by this quick analysis, in the present paper we will investigate the M=1/3 ground state of the
trimerized antiferromagnetic spin $S=\frac12$ zig-zag ladder. Specifically,
we start with a homogeneous Heisenberg antiferromagnetic
zig-zag ladder with exchanges $J_1=J_2$ at $1/3$ magnetization.
This situation is well above the critical coupling $J_2 = 0.487 J_1$\cite{Okunishi2004},
so that
the system exhibits the magnetization plateau with
a three-fold degenerate ground state.
We then consider a lattice deformation of period three,
giving rise to a modulation of the same order on NN and NNN
exchange couplings. The Hamiltonian can be written as
\beq
H=\sum_i(J_{i}\; \spin_i\cdot\spin_{i+1} + \tilde{J}_i \;\spin_i\cdot\spin_{i+2}),
\label{fullsystem}
\eeq
where $\spin_i$ denotes a spin $\frac12$ operator at site $i$.
The modulation is given by NN antiferromagnetic couplings $J_i>0$
forming a sequence of period three with $J_1=J_2=J$, $J_3=J'$,
and NNN antiferromagnetic couplings $\tilde{J}_i>0$ also
forming a sequence of period three with $\tilde{J}_1=J$, $\tilde{J}_2=\tilde{J}_3=J'$.

\begin{figure}[htbp]
    \centering
        \includegraphics[width=0.5\textwidth]{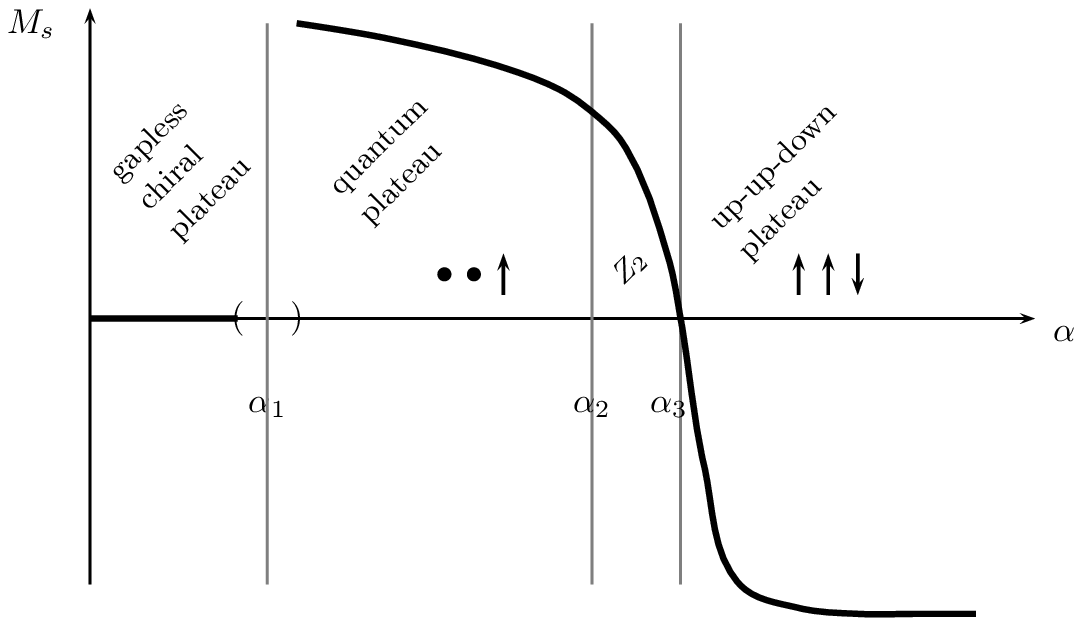}
    \caption{Schematic description of the ground state phase diagram
of the spin system shown in Fig.\ \ref{cadena}. $M_s$ is a magnetization order parameter to be referred to
in Eq.\ (\ref{orderparameter}).}
    \label{schematicPhDiag}
\end{figure}

As we will see, the interplay between frustration
and modulation gives rise to
a very rich ground state phase diagram involving a number of quantum phase transitions of different order,
including up-up-down, quantum plateau and gapless  chiral plateau states.
In summary, our analysis will show that the system adopts a unique up-up-down ground state for
$\alpha\equiv J'/J > 1$ separated
by a first order transition at $\alpha_3=1$ from a $Z_2$ degenerate ground state phase
extending to $\alpha$ slightly less than $1$. At some finite distance $\alpha_2<1$
we find a second order transition
of Ising class to a unique quantum plateau state.
These quantum phase transitions should be contrasted with the
stability of the three-fold degenerate ground state against translation invariant modifications of the
zig-zag couplings $J_1,J_2$ \cite{Okunishi2003}.
Towards the strong
frustration regime $\alpha \ll  1$ we find another transition at $0<\alpha_1< \alpha_2$ to a
gapless chiral phase with uniform magnetization ground state.
Thus we provide a
phase diagram with $J'$ starting from zero (decoupled triangular trimers) up to $J' > J$,
schematically shown in Fig.\ \ref{schematicPhDiag}.
The analysis relies on bosonization, renormalization group and block-spin perturbations, supported by
numerical diagonalization on finite size systems.
A detailed knowledge of these transitions will
ultimately help understanding the possibility of spin-Peierls-like phenomenon in magnetized materials.

The paper is organized as follows.
In section II we study the regime of weak modulation $J' \sim J$ by bosonization, obtaining a triple sine-Gordon
effective theory. In section III we study the
strong frustration regime, where block-spin perturbation theory is applicable.
Effective Hamiltonians at first and second order are analyzed. In section IV we
study numerically the ground state and low lying excitations in finite systems by exact diagonalization;
this supports the semi-quantitative bosonization results and provides a bridge between weak modulation and
strong frustration regimes. In section V we present our conclusions and prospects for future work.

\section{Weakly modulated systems}

We first analyze the stability of the zig-zag ladder plateau ground state at $M=1/3$ against
small modulated perturbations, as defined in Eq.\ (\ref{fullsystem}) for $J' \sim J$. The microscopic
Hamiltonian for the system can be conveniently written as
\beqa
    H &=& \sum_i
    \left[
    (J-\frac{\epsilon}{2})\spin_i \cdot \spin_{i+1}
    +(J-\epsilon)\spin_i \cdot \spin_{i+2}
    -h S^z_i
    \right]
    \nn \\
    &&
    -\epsilon \sum_i \cos(i\frac{2\pi}{3})\spin_i \cdot \spin_{i+1}
    \nn \\
    &&
    +\epsilon \sum_i \cos((i-1)\frac{2\pi}{3})\spin_i \cdot \spin_{i+2},
\label{hom+osc}
 \eeqa
where $\epsilon = 2(J-J')/3$. The first line describes a homogeneous zig-zag ladder with
$J_1=J-\epsilon/2$ and $J_2 = J-\epsilon$ in
an external magnetic field $h$, while the rest is a modulated perturbation of
period three.

Once the magnetic field is tuned to the plateau range,
the homogeneous part is well described \cite{Lecheminant2004,Hida2005} by bosonization with
an effective massive sine-Gordon Hamiltonian
\beq
H_{0} =
\frac{v}{2}\int dx \left[ \frac{1}{K}(\partial_x\phi)^2+K\,(\partial_x\tilde{\phi})^2\right]
-g \cos[3\sqrt{4\pi}\phi], \label{Hhom}
\eeq
where $\phi$ is a real bosonic field with spin wave velocity $v$ and compactification
radius $R=1/\sqrt{4\pi}$,
\footnote{In our conventions, this ensures that the bosonized spin operators are single-valued,
as the compactification radius sets $\phi \equiv \phi+2\pi R$. See for instance [\onlinecite{CabraPujol}].}
and $\tilde\phi$ is its dual field defined by
$\partial_x\tilde\phi = \partial_t\phi$. The Luttinger parameter $K$ takes into account
renormalization due to spin interactions.
The presence of the third harmonic of the bosonic field arises from a triple Umklapp term,
only allowed at the Fermi momentum $k_F=\pi/3$ fixed by the magnetization M=1/3. Moreover,
$K$ has to be sufficiently small, $K<2/9$, rendering this term
a relevant conformal perturbation, and $g$ has to be positive, in order to fit
numerical results.

\begin{figure}[htbp]
    \centering
        \includegraphics[width=0.5\textwidth]{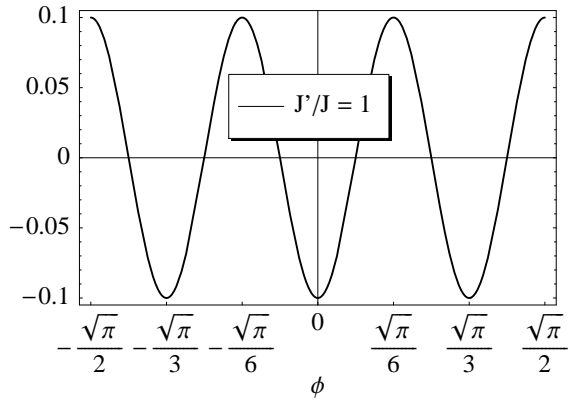}
    \caption{Semiclassical potential for the homogeneous system $J=J'$ (in arbitrary units).
    There are three minima in the compactified range $(-\frac{\sqrt{\pi}}{2},\frac{\sqrt{\pi}}{2})$,
    corresponding to the three non-equivalent up-up-down states.}
    \label{simple}
\end{figure}

Indeed, in a semiclassical analysis, the relevant third harmonic $-g \cos[3\sqrt{4\pi}\phi] $
is considered as a potential; it has three non equivalent
minima for the compactified boson field $\phi$ (see Fig.\ \ref{simple}) signaling a
three-fold degenerate ground state.
The mapping to spin variables, following the usual rule for magnetized systems \cite{CabraPujol},
shows that each of the minima for $\phi$ corresponds to the so called ``classical plateau states'' \cite{Hida2005},
namely up-up-down states related by lattice translations $i \to i+1, i+2$.
This is just what numerical results\cite{Okunishi2003} show, a three-fold degenerate ground state with
spontaneous $Z_3$ symmetry breaking to states with up-up-down local magnetization
structure. The denomination of ``classical plateau states'' indicates that these are essentially the states
obtained in the Ising limit of the present spin system.
Interestingly, this description is stable for a wide range of homogeneous variations of
$J_1$ and $J_2$ couplings.

The second and third lines in Eq.\ (\ref{hom+osc}) represent a modulated perturbation to the
homogeneous zig-zag ladder described above.
The key point in our presentation is that, after bosonization, these perturbations
provide first and second harmonics in the Hamiltonian that are commensurate with the short scale oscillations
given by $k_F=\pi/3$.
Specifically, these terms yield an effective contribution which can be
recasted as
\beq
H_{mod} \sim
\epsilon \int dx \left[ C\; \cos[\sqrt{4\pi}\phi]-\cos[2\sqrt{4\pi}\phi]\right],
\label{Hmod}
\eeq
where $C$ is a coefficient of order 3 ($C=1+\frac{2\pi}{3}$). These harmonics have smaller scaling
dimension than the third one, thus providing relevant perturbations that will compete with it. The
presence of the first harmonic substantially modifies the theory. It is now the leading
perturbation, and the effective field theory $H_{0} + H_{mod}$ is the so called triple sine-Gordon
model \cite{HoghJensen1982}. Extensive analysis of competition between harmonics has been
presented in [\onlinecite{Delfino1997,Nersesyan2000,Takacs2001}], mainly focused on the double sine-Gordon
model. The three-frequency case has been recently discussed in detail in [\onlinecite{Toth2004}].

Regarding spin systems, a model similar to ours has been recently analyzed by Hida and Affleck in
[\onlinecite{Hida2005}], where
a modulated perturbation on NN couplings of a zig-zag ladder was proposed as a way to diminish
frustration and enforce singlet formation on the M=1/3 plateau ground state.
The effective theory obtained by these authors contains first and third harmonics of the bosonic field.
In the present case, the second harmonics makes necessary an extension of the above mentioned results.

We will first perform the semiclassical analysis of the ground state,
using the bare coefficients in Eq.\ (\ref{Hmod}).
Thereafter, the Renormalization Group (RG) flow of the relevant perturbations will be discussed.

\begin{figure}[htbp]
    \centering
        \includegraphics[width=0.5\textwidth]{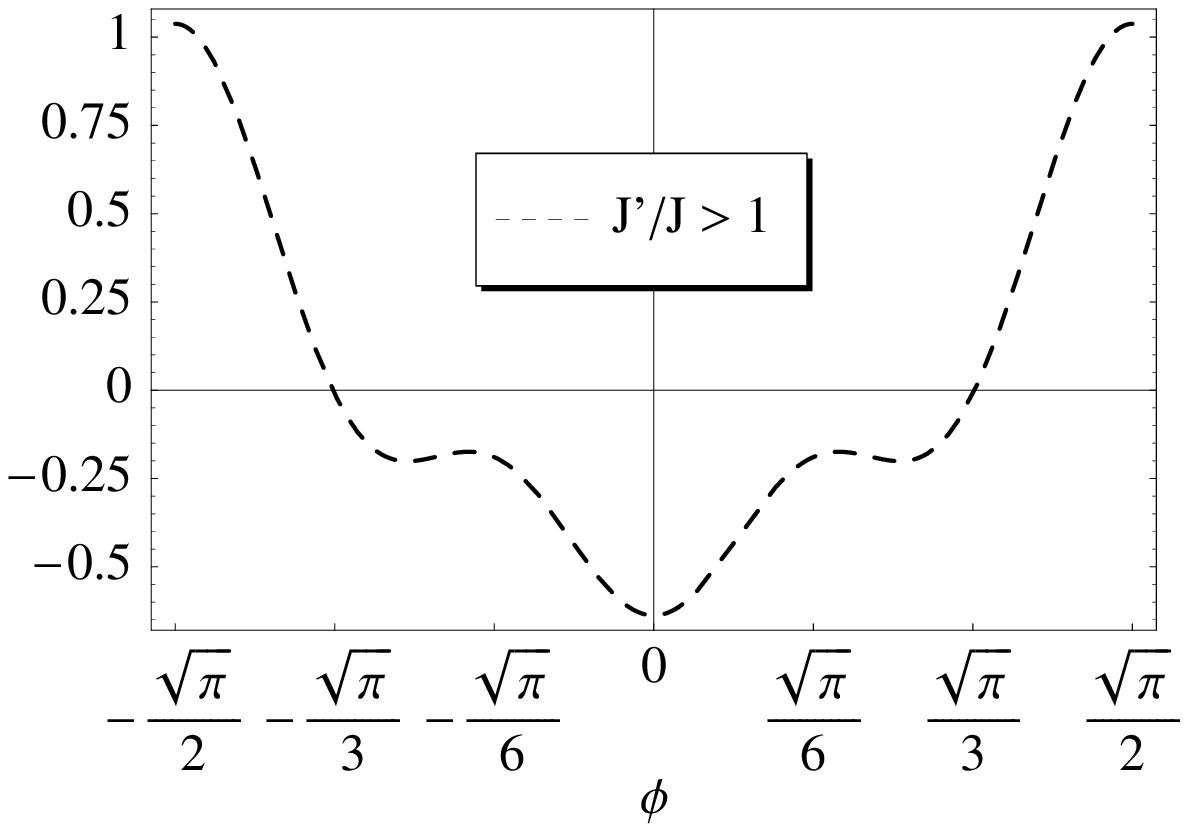}
        \includegraphics[width=0.5\textwidth]{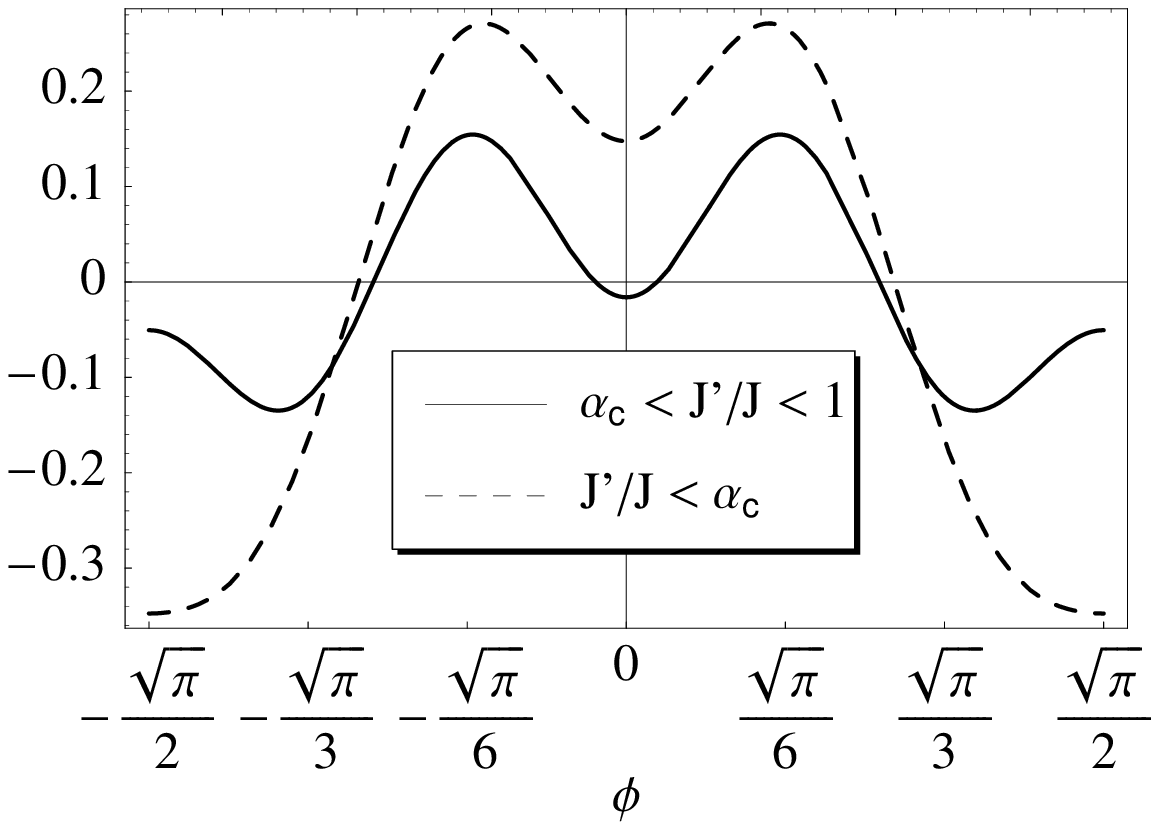}
    \caption{Modification of the semiclassical potential by relevant harmonics (in arbitrary units).
    For $J' > J$ (upper panel), the single central minimum pins the system in a particular up-up-down state.
    For $J' < J$ (lower panel), the minima structure takes the ground state first to double degeneracy and then,
    at some finite critical value of $J'/J$, to a
    single minimum shifted $ \sqrt{\pi} /6$ with respect to non-perturbed position.
    This shift selects a quantum plateau state out of the triple degenerate up-up-down states.}
    \label{potential-plot}
\end{figure}

For $J' > J$ the basic harmonic dominates the minimum of the potential
(see Fig.\ \ref{potential-plot}, upper panel),
so that the perturbation selects only one configuration $\phi = 0$
corresponding to one of the up-up-down states
out of the three-fold degenerate ground state
(the one with  spins down on sites numbered by $3i+2$ in Fig.\ \ref{cadena}).

A richer scenario occurs for $J' < J$ when the first and second
harmonics are in conflict with the triple harmonic term (see Fig.\
\ref{potential-plot}, lower panel).
In this case triple degeneracy is transformed first into double by an infinitesimal
perturbation. From the three degenerate states at $J'=J$
it is precisely the state selected for $J'>J$ the one that now becomes excited,
while the other two potential minima are selected and shifted
from their commensurate up-up-down positions. This phase is characterized by
$Z_2$ reflection symmetry, spontaneously broken by each of the potential minima.
When $J -J'$ reaches some \emph{finite} positive value these two minima collapse
into a single one at $\phi=\sqrt{\pi}/2 \equiv - \sqrt{\pi}/2$,
lifting the degeneracy completely. Moreover, the position of the minimum
is shifted to a commensurate field configuration which was a maximum for the homogeneous ladder.
In terms of spins, the system is pinned in a very
different configuration, the so called ``quantum plateau state'' \cite{Hida2005}.
This configuration is characterized by spin singlets alternating with spin up
sites (in our case, with spin up at sites numbered by $3i+2$ on Fig.\ \ref{cadena}). Given the
symmetries and degeneracy of the ground states on both sides of the critical point, one may
conjecture that the transition belongs to the second order Ising universality class
\cite{Delfino1997}.

\begin{figure}[htbp]
    \centering
        \includegraphics[width=0.4\textwidth]{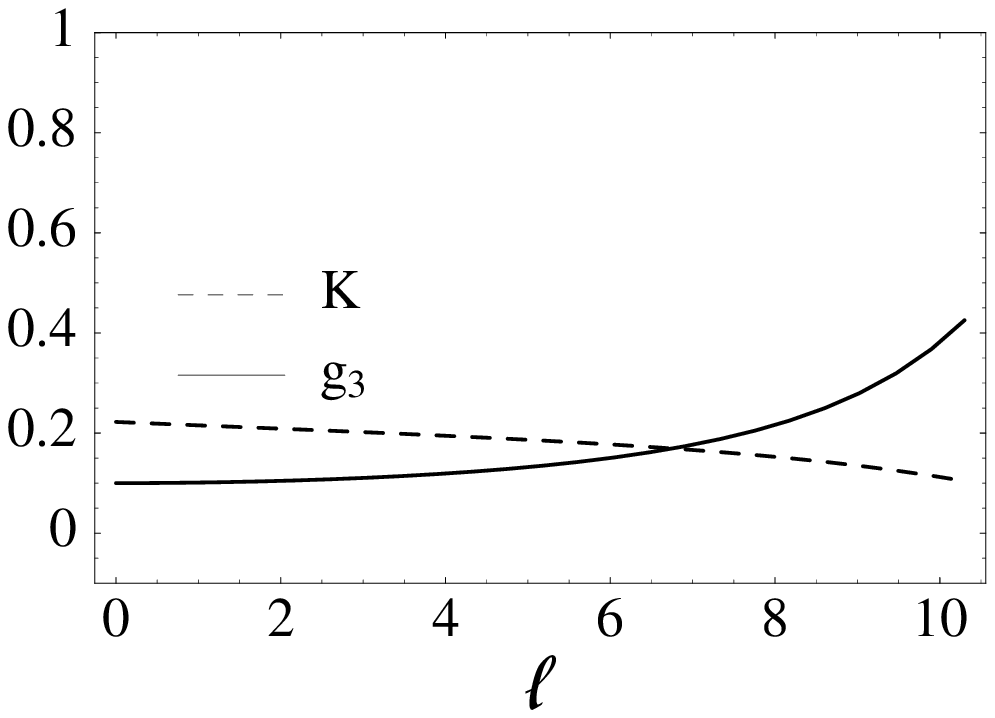}
        \includegraphics[width=0.43\textwidth]{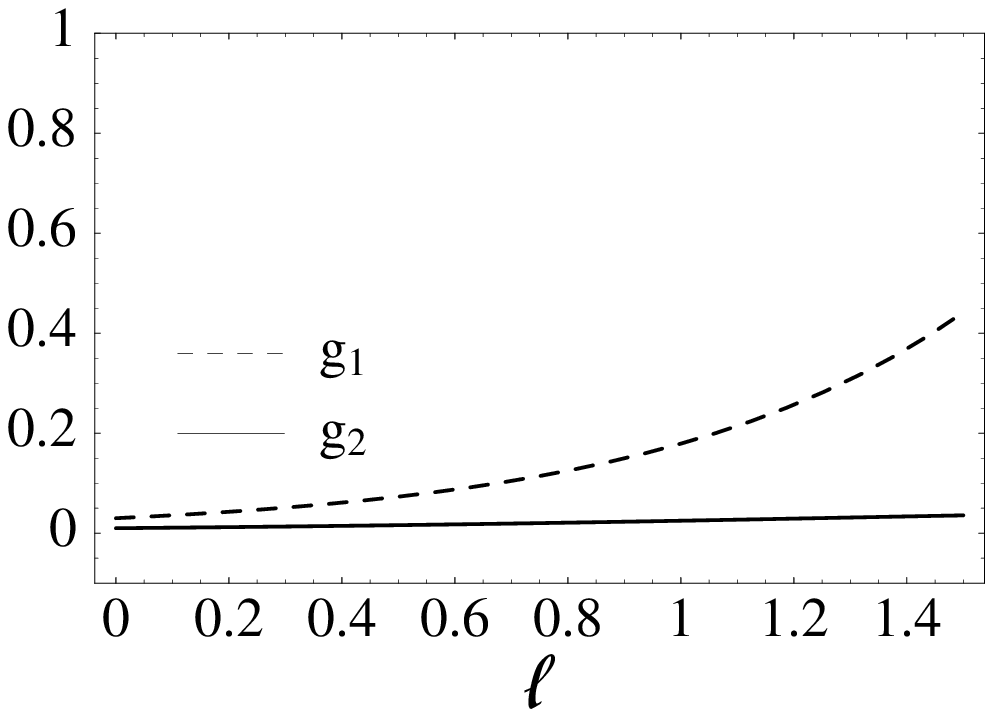}
    \caption{Representative RG flow of triple sine-Gordon couplings in terms of the length scale.
    The region of interest is given by initial conditions
    $K<\frac{2}{9}$, $g_3\gg g_2,g_1$  and $g_1 \simeq 3 g_2$, corresponding to the bare Hamiltonian in Eqs.\
    (\ref{Hhom}, \ref{Hmod}).
    (a) Couplings $K$ and $g_3$ describing the homogeneous zig-zag ladder.
    (b) First and second harmonics couplings $g_1$ and $g_2$ associated to modulated perturbations.
    }
    \label{RG_flow}
\end{figure}

The preceding discussion describes the competition between harmonics,
depending on the signs and values of their bare coefficients. The results are particularly
sensitive to the relative coefficients of first and second harmonics.
In order to describe the long distance effective theory, the RG flow of the coefficients
must be analyzed.
To this aim, let us write the triple sine-Gordon Hamiltonian in a compact form
\begin{eqnarray}
H &=& H_{LL}+ g_1\,\cos[\sqrt{4\pi}\phi]-g_2\,\cos[2\sqrt{4\pi}\phi]-\nn\\
&&-g_3\,\cos[3\sqrt{4\pi}\phi]
\end{eqnarray}
where $H_{LL}$ is the free boson Hamiltonian (Luttinger liquid) with Luttinger parameter $K$.
All the  present cosine terms are  relevant perturbations when $K<\frac{2}{9}$, and the signs
have been chosen so as to represent the bare situation with positive coupling constants $g_1,g_2,g_3$.
The region of parameters we are interested in corresponds to $g_3$ of order of unity,
describing the homogeneous ladder $Z_3$ symmetric ground state,
and small $g_1,g_2 \propto \epsilon$ describing the modulation effects.
Up to second order in $g_1$, $g_2$ and $g_3$, the perturbative RG equations read
\begin{eqnarray}
\label{eqs_RG}
\frac{d}{dl}\frac{1}{K}&=&\frac{9\pi}{2}\,g^2_3\nn\\
\frac{dg_3}{dl}&=&\frac{9}{8\pi}(2-9 K)g_3\nn\\
\frac{dg_1}{dl}&=&(2-K)g_1+\frac{13}{18}\,g_1g_2-\frac{5}{18}g_2g_3\\
\frac{dg_2}{dl}&=&(2-4 K)g_2-\frac{4}{9}\,g_1g_3\nn,
\end{eqnarray}
where $l$ is the length scale.
The first two equations are the well-known  sine-Gordon RG equations \cite{Giamarchi};
as expected, $g_3$ flows towards strong coupling and $K$ decreases.
The other two equations can then be solved
using the solutions for $K,g_3$.
We performed a numerical analysis of Eqs.\ (\ref{eqs_RG}) in the region of interest. The typical form of the
RG flow is shown in Fig.\ \ref{RG_flow}.
The salient features are that $g_1$ remains positive and grows faster than $g_3$, allowing for the
modification of the semiclassical potential, and that
in the perturbative regime $g_2$ remains positive and smaller than $g_1$.
This analysis proves that the phase transitions depicted above are correct in
the low energy (long distance) renormalized effective theory.

We have then found that the bosonization analysis predicts a first order phase transition at $J'=J$,
as the configurations selected
for ground states by the classical potential jump between different minima of the potential,
and a second order phase transition of Ising class at some finite value of $\alpha_2 = J'/J$,
where two degenerate minima merge continuously into a single one.

A natural question is now if the quantum plateau phase remains stable for $J'/J<\alpha_2$.
It is easy to see that there should be a further phase transition, as for $J'/J \ll 1$ the system
is described by almost decoupled triangular trimers. We investigate this phase in the following section.

\section{Strong Frustration}

In order to analyze the ground state of the system beyond the weak modulation regime,
we choose in this section a different starting point, namely the strong frustration regime. We consider
the system as consisting of frustrated triangles of
spin $S=\frac12$ with antiferromagnetic exchange coupling $J$,
weakly coupled among them with exchange coupling $J' \ll J$.
\begin{figure}[htbp]
    \centering
        \includegraphics[width=0.35\textwidth]{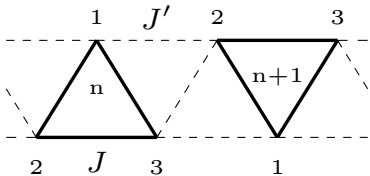}
    \caption{Schematic description of the spin system in the strong frustration regime, with
    $J' \ll J$ as indicated by bold line trimers.}
    \label{cadena_tri}
\end{figure}
In this regime, the system can be thought of as a quasi
one-dimensional weakly coupled chain of triangles.
The topology of weak couplings has been chosen so as to recover the
zig-zag chain in the limit $J'/J \to 1$, but similar systems with more symmetric inter-triangle couplings
have been considered in relation with spin tubes \cite{Fouet2005}.
Notice that the M=1/3 plateau is very robust for $J' \ll J$,
as becomes apparent in the limit case $J'=0$; we
can then safely consider a plateau regime with M=1/3 even for small magnetic fields.

For convenience, in this section we enumerate the triangular trimers with an index $n$ and
rename spins as $\spin_n^a$ ($a=1,2,3$ inside each trimer, see Fig.\ \ref{cadena_tri})
to write the Hamiltonian in Eq.\ (\ref{fullsystem})
as
\beqa
\lefteqn{H=\sum_{n} J\left( \spin_{n}^1\cdot \spin_{n}^2 + \spin_{n}^2\cdot \spin_{n}^3 +
\spin_{n}^3\cdot \spin_{n}^1
\right)} \label{system} \\
&+& \sum_{n} J'\left( \spin_{n}^1\cdot \spin_{n+1}^2 + \spin_{n}^3\cdot \spin_{n+1}^1 + \spin_{n}^3\cdot
\spin_{n+1}^2 \right).
\nn
\eeqa

We perform a systematic block-spin perturbative analysis \cite{Subrahmanyam1995,Mila1998}
around the highly degenerate exact ground state of the system of
decoupled trimers ($J'=0$) at zero magnetic field.
For low energy, the  spin operators at each vertex can be factorized in terms of the triangle total spin
operator and a pseudo-spin chiral operator  \cite{Raghu2000} as
\beq
\spin_n^a = \frac{2}{3} \spin_{n,T} (\frac{1}{2} - T_n^a),
\label{spin-representation}
\eeq
where
\beq
\spin_{n,T}=\spin_{n}^1+\spin_{n}^2+\spin_{n}^3
\label{total-spin}
\eeq
is the total spin operator of the $n$-th triangle projected onto the $S=\frac12$ low energy sector.
Operators $T_n^a$  act on a chiral sector as
\beqa
T_n^1 &=& \tau_n^+ + \tau_n^- \, , \nn\\
T_n^2 &=& \omega^2\tau_n^+ + \omega\tau_n^- \, , \nn\\
T_n^3 &=& \omega\tau_n^+ + \omega^2\tau_n^- \, ,
\label{Ts}
\eeqa
where
$\vec{\tau}_n$ are generators of the $S=\frac12$ pseudo-spin sector,
with
$\tau_n^z=\frac{2}{\sqrt{3}}\spin_{n}^1 \cdot \spin_{n}^2 \wedge \spin_{n}^3$,
and $\omega = e^{i\, 2\pi/3}$ a primitive cubic root of unity.

In order to study the M=1/3 plateau ground state, we consider the system under the action of an
external magnetic field $h^*$ high enough so as to
force total magnetization $1/3$, but still low enough to discard excited spin states $S=\frac32$ at any
triangle. This regime is clearly available for small $J'/J$.
The trimer spin degree of freedom is then
saturated to $S_{n,T}^z=\frac12$, leaving a pseudo-spin chain Hamiltonian describing non magnetic excitations.
Notice that the magnetic field couples only to the total spin and plays no role in the pseudo-spin sector.

At first order in block-spin perturbations we get a simple XY-like nearest trimers Hamiltonian,
\beq
H_{eff}^{(1)}=\frac{\sqrt{3}}{9}J'\sum_n
\big(e^{-i\frac{\pi}{6}}\tau^+_n \tau^-_{n+1} +\ e^{i\frac{\pi}{6}}\tau^+_{n+1}\tau^-_n\big),
\label{Heff1}
\eeq
where we have dropped constant terms, including the magnetic field.
One can perform a gauge transformation
\beqa
\tau^{+}_{n}& \rightarrow & e^{-i\, n \frac{\pi}{6}}\,\tau^{+}_{n},\nonumber\\
\tau^{z}_{n}& \rightarrow & \tau^{z}_{n},
\label{gauge}
\eeqa
that preserves $SU(2)$ commutation relations, to eliminate phases. Then the usual Jordan-Wigner
transformation maps non magnetic chiral excitations onto a gapless theory of free spinless fermions.
The ground state is thus described by a half-filled fermion band.

From this effective model, one can also analyze the non magnetic excitations above the plateau ground state,
which form a gapless continuum with non degenerate ground state.
Using Eqs.\ (\ref{total-spin},\ref{Ts},\ref{gauge}) one can evaluate
the ground state expectation value and correlations of spin operators.
The spin density in the direction parallel to the applied field is uniform and simply equals
average magnetization,
\beq
\langle S_n^{z\,a} \rangle =1/6,
\label{spin-density}
\eeq
whereas in-plane averages vanish.
Equal time spin-spin correlation function in the direction parallel to the applied field behaves as:
\beq
\langle S_{n}^{z\, a} S_m^{z \,b}\rangle =\frac{1}{36} +\frac{{\mathrm {const}}
}{\sqrt{n-m}}\cos((n-m)\frac{\pi}{6}+\frac{2\pi}{3}(a-b)),
\label{spin-correlation}
\eeq
with $a,b=1,2,3$
and thus
show long range order (each triangle being in a spin $+1/2$ state), plus algebraically decaying oscillations.
The in-plane-$XY$ correlation functions decay exponentially.
This picture is a consequence of the fact that non magnetic excitations are gapless,
while magnetic excitations are gapped.

From the above discussion one can state that for $J' \ll J$
the spin system at $M=1/3$ presents a gapless phase corresponding to non magnetic chiral degrees of freedom
described by a Luttinger liquid with $K=1$.

In order to look for a quantum phase transition towards the quantum plateau state, it is necessary to
construct the block-spin effective Hamiltonian at second order.
This derivation requires a much longer calculation, and provides next nearest trimers interactions.
Skipping details, once total spin is saturated to $\frac12$ at each triangle
we get for the second order correction
\beqa
H^{(2)}_{eff}/J &=&
\alpha^2 \left\{\frac{2}{27}\sum_n[\tau^{+}_{n}+\tau^{-}_{n}]+\right.\label{Heff2} \\
&&+\frac{5}{162}\sum_n\,
[e^{-i\frac{2\pi}{3}}\tau^{+}_{n}\tau^{-}_{n+1}+e^{i\frac{2\pi}{3}}\tau^{+}_{n+1}\tau^{-}_{n}]-\nn\\
&&-\frac{1}{3}\sum_n\tau^{z}_{n}\tau^{z}_{n+1}+\nn\\
&&+\frac{2}{81}\sum_n[e^{i\frac{2\pi}{3}}\tau^{+}_{n}\tau^{-}_{n+2}+
e^{-i\frac{2\pi}{3}}\tau^{+}_{n+2}\tau^{-}_{n}]-\nn\\
&&-\frac{2}{27}\sum_n[\tau^{+}_{n}\tau^{+}_{n+1}+\tau^{-}_{n+1}\tau^{-}_{n}]-\nn\\
&&\left.
-\frac{4}{81}\sum_n[\tau^{+}_{n}\tau^{+}_{n+1}\tau^{+}_{n+2}
+\tau^{-}_{n+2}\tau^{-}_{n+1}\tau^{-}_{n}]\right\},\nn
\eeqa
where again we have dropped constant contributions.
Terms in the second line are similar to those appearing at first order in Eq.\ (\ref{Heff1}), so we
propose a different gauge transformation in order to eliminate the phase in XY terms
\beqa
\tau^{+}_{n}&\rightarrow&e^{i\lambda(\alpha)n}\,\tau^{+}_{n},\nonumber\\
\tau^{z}_{n}&\rightarrow&\tau^{z}_{n},
\label{second_gauge}
\eeqa
where $\lambda(\alpha)=\arctan[\frac{\sqrt{3}(5\alpha-18)}{5\alpha+54}]$, rendering
the total effective Hamiltonian written as
\beqa
\lefteqn{H_{eff}/J=} \nn\\
&&-\rho(\alpha)\sum_n\,[\frac{1}{2}(\tau^{+}_{n}\tau^{-}_{n+1}+\tau^{+}_{n+1}\tau^{-}_{n})+
\Delta^z(\alpha)\tau^{z}_{n}\tau^{z}_{n+1}]+\nn\\
&&+\frac{2}{81}\alpha^2\sum_n[e^{-i2(\lambda_{(\alpha)}-\frac{\pi}{3})}\tau^{+}_{n}\tau^{-}_{n+2}+h.c.]+\nn\\
&&+\frac{2}{27}\alpha^2\sum_n[e^{i\lambda_{(\alpha)}n}\tau^{+}_{n}+h.c.]-\nn\\
&&-\frac{2}{27}\alpha^2\sum_n[e^{i\lambda_{(\alpha)}(2n+1)}\tau^{+}_{n}\tau^{+}_{n+1}+h.c.]-\nn\\
&&-\frac{4}{81}\alpha^2\sum_n[e^{i\,3\,\lambda_{(\alpha)}(n+1)}\tau^{+}_{n}\tau^{+}_{n+1}\tau^{+}_{n+2}+h.c.],
\label{Htotal_chis}
\eeqa
where  $\Delta^z(\alpha)=\frac{\alpha^2}{3\rho(\alpha)}$ and
$\rho(\alpha)=\alpha/81\sqrt{972+25\alpha^2}$.

In Eq.\ (\ref{Htotal_chis}) the second and third lines represent an extended Heisenberg $XXZ$ model with
NN and NNN interactions for pseudo-spin operators. NN interactions have acquired the anisotropy parameter
$\Delta^z(\alpha)$,
while NNN interactions are XY-like, with a shifted band (because of the
phase in the couplings);
inspection of the coefficients shows that $0< \Delta^z(\alpha) < 1$ and NNN couplings are small compared with
NN ones.
All the other terms include phases that depend on the position,
and will generally cancel out in the continuum limit,
so that the continuous $U(1)$ symmetry broken by some of them is recovered.
The low energy behaviour of this pseudo-spin $\frac12$ model is then appropriately described by bosonization,
where it is most clearly seen that oscillatory terms are incommensurate
and can be neglected.
We are essentially left with a one-dimensional spin $S=\frac{1}{2}$ anisotropic
Heisenberg model.

Applying the standard bosonization procedure we derive the low energy Hamiltonian
\begin{eqnarray}
H_{eff}& \approx &\frac{v}{2}\int dx\big[\frac{1}{K}(\partial_x\varphi)^2+ K(\partial_x\tilde{\varphi})^2\big]
\nn\\
&&-\Gamma\int dx\, \cos[2\sqrt{4\pi}\varphi],
\label{Hboson}
\end{eqnarray}
where the Fermi velocity, the Luttinger parameter and the coupling $\Gamma$ depend on $J'/J$.
For small $J'/J$ the harmonic perturbation is strongly irrelevant, so that the effective theory still describes
a gapless phase.
As $J'/J$ increases, the conformal dimension of the harmonic perturbation decreases and one expects
from Eq.\ (\ref{Hboson})
a second order Brezinskii-Kosterlitz-Thouless (BKT) like phase transition to a massive phase.
A detailed computation allows to estimate a critical value $J'/J \sim 0.5$, a value that is
possible beyond the validity of block-spin perturbation theory.

\section{Numerical analysis}

In order to support the analytical results in the preceding sections, and to explore the intermediate regime
of $J'/J$ not covered by perturbations around $J'=0$ and $J'=J$,
we performed exact diagonalization analysis
on finite systems with 12, 18 and 24 spins using periodic
boundary conditions. However, in attempting to avoid misleading
results in the approximants and extrapolations referred to below,
we discarded the 12 spin data.

\begin{figure}[htbp]
    \centering
        \includegraphics[width=0.5\textwidth]{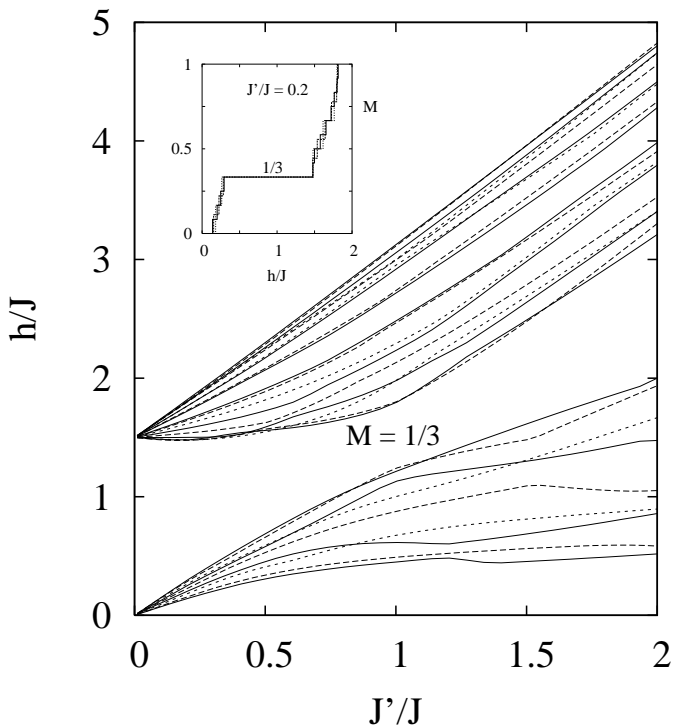}
    \caption{Magnetic phase diagram obtained by exact diagonalization of finite system with
    12 (doted line), 18 (dashed line) and 24 spins (solid line).
    Inset: a typical magnetization curve.}
    \label{phasediagram}
\end{figure}

We have first confirmed the presence of the $M=1/3$ magnetization plateau, all over the range from
$J'=0$ to $J' = 2J$. The magnetic phase diagram showing the magnetic field $h$
necessary for level crossing between
available magnetizations in several finite size systems is presented in Fig.\ \ref{phasediagram}.
A noticeable finite increment in $h$ separates magnetic excitations from the $M=1/3$ ground state,
indicating the magnetization plateau.
This is interpreted to remain in the infinite size scaling limit.

\begin{figure}[htbp]
    \centering
        \includegraphics[width=0.45\textwidth]{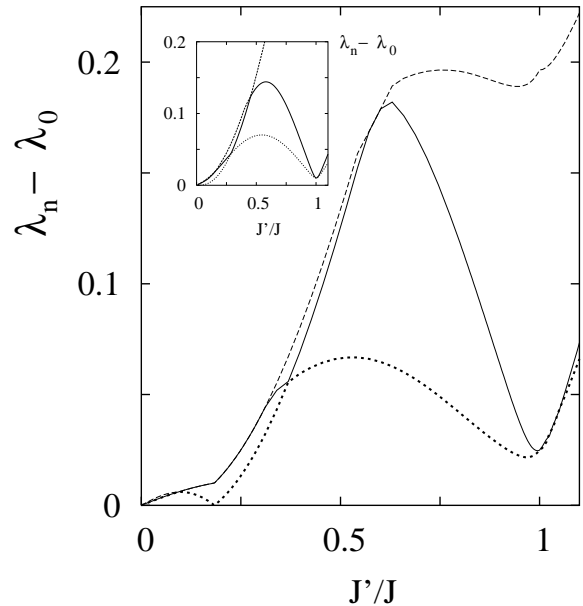}
    \caption{Non magnetic excitations $\lambda_n$ for 24 spins with periodic
boundary conditions above the $M = 1/3$ ground state energy
$\lambda_0$, for $n= 1$ (dotted line), 2 (solid line), and 3 (dashed line).
The inset shows the corresponding results for 18 spins.
Numerical data in all subsequent figures were also obtained fixing
$M = 1/3$.
}
    \label{3priexci}
\end{figure}

We have then computed the ground state and first three excitation energies in
the subspace of magnetization $M=1/3$ in
a wide range of couplings $0< J'/J< 1.5$.
In Fig.\ \ref{3priexci} we plot
the gaps to excited energies in terms of $J'/J$.
The triple degeneracy of the zig-zag ladder ground state is, within finite size effects,
qualitatively observed at point $J'/J=1$ in agreement with
Ref.\ [\onlinecite{Okunishi2003}].
Moreover, assuming that the
ground state will become degenerate in the thermodynamic limit, our numerical data
are compatible with the picture that the triple degeneracy is lifted to a unique ground state for
$J'/J>1$. It also seems to be partially lifted to double degeneracy
for $J'/J \lesssim 1$ where one level rapidly separates
but another gets closer to the ground state, and only then raises
lifting the remaining degeneracy.

One can estimate the locations of the Ising and BKT transition critical points
mentioned in sections II and III
by considering the Callan-Symanzik  $\beta$-function
developed in the context of phenomenological renormalization group
by Roomany and Wyld [\onlinecite{RW}]. This technique
can handle situations in which the phase transition is not necessarily
characterized by a power law decay of the spectrum gap, {\em i.e.} an ordinary
second order transition,  but also by a singular BKT form. In the former
case the $\beta$-function exhibits a simple zero whereas in the latter
situation it vanishes with an algebraic singularity.
This function can be estimated from finite lattice data  by the Roomany-Wyld (RW)
approximant, which in our notation reads
\begin{eqnarray}
\lefteqn{\beta_{\rm RW} (\alpha) =
\alpha\, \ln \left[\, \frac{ N+6}{N} \; \frac{\Delta_{N+6} (\alpha)}{ \Delta_N (\alpha)}\,\right] / }  \\
&&\ln \left( \frac{N+6}{N} \right) \left\{\,1 \,+\,\frac{1}{2} \,\alpha \,\partial_\alpha \,
\ln \left[  (N+6) \,\Delta_{N+6} (\alpha)  \, N \,\Delta_N (\alpha)\,\right] \,  \right\} \,,
\nonumber
\label{RW}
\end{eqnarray}
where $\Delta_N$ is the spectrum gap per spin.
Notice that whenever the phenomenological renormalization
condition $(N+6) \Delta_{N+6}(\alpha) = N \Delta_N(\alpha)$
is satisfied, the $\beta$-function shows a zero. Furthermore, it
behaves as $\beta (\alpha) \sim (\alpha - \alpha_c) /\nu$, from where the slope
at $\alpha_c$ is related to the exponent of the second order transition.
Instead, in the vicinity of a singular transition of the form
$\Delta \propto \exp -(\alpha-\alpha_c)^{-\sigma}$, one has
$\beta \sim (\alpha-\alpha_c)^{1+\sigma}$ which ultimately determines  both $\alpha_c$
and  $\sigma$.
\begin{figure}[htbp]
    \centering
        \includegraphics[width=0.4\textwidth]{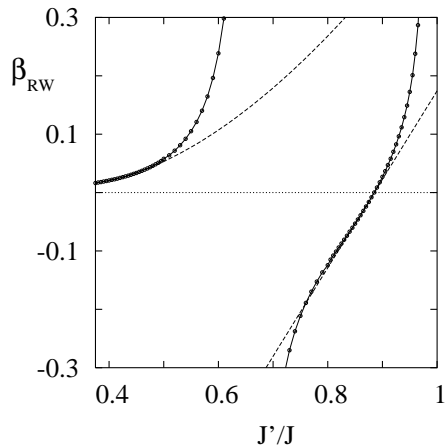}
    \caption{The Roomany - Wyld approximant ($N = 24, 18$) of the Callan-Symanzik
$\beta$-function (circles). The left and right curves are respectively consistent
with a BKT phase transition at $\alpha_1 \approx 0.35$ and with an Ising
transition at $\alpha_2 \approx 0.88$.
Dashed curves fit the numerical data
with the parameters referred to in the text.}
    \label{beta}
\end{figure}
In Fig.\ \ref{beta} we plot the RW approximant computed from 18 and 24 spins data. The leftmost
curve characterizes a singular transition with $\alpha_1 \sim 0.35$ and $\sigma \sim 0.7$ (though
not exhibiting a strict zero, possibly an artifact of our small lattice sizes) compatible with the BKT type,
while the rightmost one
typifies a conventional transition at $\alpha_2 \sim 0.88$ with $\nu \sim 0.67$ consistent with the Ising class.

\begin{figure}[htbp]
    \centering
        \includegraphics[width=0.45\textwidth]{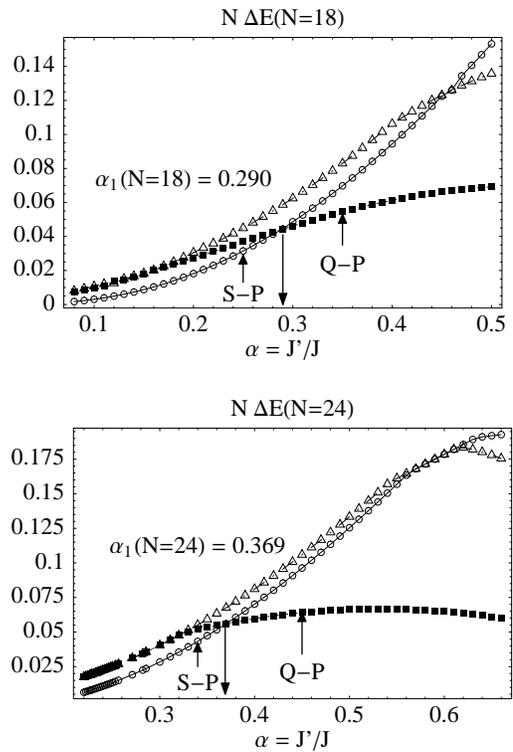}
   \caption{Detail of scaled excited levels crossing used in the level spectroscopy analysis.}
\label{LS method}
\end{figure}

The BKT transition can also be estimated from level crossing spectroscopy
\cite{LS1, LS2, LS3} of the low-lying states with different symmetries.
In Fig.\ \ref{LS method}, we show the size-scaled excitation energies $N \Delta E(N)$ as functions of $J'/J$
for finite-size clusters with $N=18$ and $N=24$ spins.
The intersection between the first and second excitations
is interpreted \cite{LS2} as the chiral fluid-quantum plateau transition critical point $\alpha_1(N)$ and
occurs at $\alpha=0.29$ and $\alpha=0.36$ respectively.
The finite size scaling of $\alpha_1(N)$ is expected to follow \cite{LS1}
\begin{eqnarray}
\alpha_1(N)&=&\alpha_1(\infty)+const \times N^{-2},
\label{alpha_c}
\end{eqnarray}
suggesting a crude extrapolation to
$\alpha_1(\infty) \approx 0.47$.

These numerical estimates are thus consistent with the existence of a BKT transition
at $\alpha_1$ in the range $0.3-0.5$ and an Ising transition at $\alpha_2$ around $0.9$.
However, a level crossing, not predicted by the analytical treatment,
is seen in Fig.\ \ref{3priexci} at $J'/J \approx 0.18$ in the 24 spins system but not in the 18 spins one.
Its presence should be checked in larger systems,
not currently available to us, as it may well be the consequence
of highly oscillating terms in Eq.\ (\ref{Htotal_chis}) which become particularly dominant
for small lattice sizes.
If confirmed, it would indicate that the BKT transition separating the gapless phase described in the strong
frustration regime from the quantum plateau described in the weak modulation regime
could be replaced by a first order one in the thermodynamic limit.
We can not make a definite
distinction from our current numerical data.

\begin{figure}[htbp]
    \centering
\includegraphics[width=0.4\textwidth]{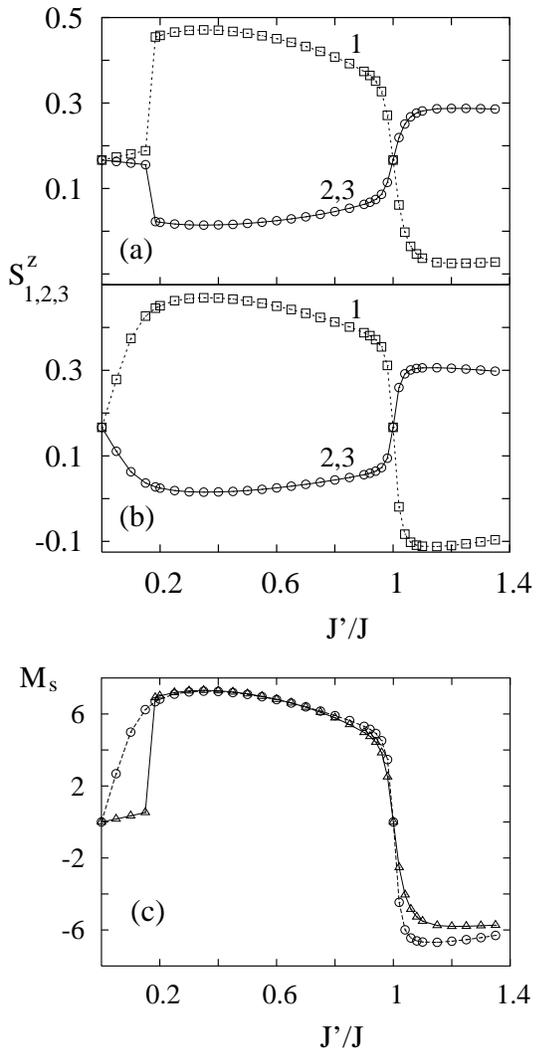}
    \caption{Upper panel (a): local magnetization of the ground state of 24 spins
as a function of the coupling parameter. Squares (circles) correspond
to middle 1 (ending 2, 3) sites in trimers. Panel (b) displays respectively
these results for 18 spins (here, the uniform phase is absent).
Lower panel (c): order parameter in Eq.\ (\ref{orderparameter}) describing modulated
magnetizations for 18 (circles) and 24 (triangles) spins. From left
to right it exhibits uniform magnetization, quantum plateau
and up-up-down ground states.
}
    \label{profile}
\end{figure}

In order to characterize the phases separated by the mentioned transition points,
we also computed the local magnetization profile for the ground state.
We found three different periodic phases according to the generic
$J'/J$ values explored.
As the profile is periodic, we report the configuration of a generic trimer,
labelling sites $a=1,2,3$ in accordance with Fig.\ \ref{cadena_tri}.
A plot with local magnetization of the ground state for $J'/J$ up to 1.4 is shown in
Fig.\ \ref{profile} (upper panel).
Notice that $a=2,3$ sites show the same magnetization.

In the strong frustration limit we observe for the 24 spins system an almost
uniform magnetized ground state for $ 0 < J'/J \lesssim 0.18$
(for instance $S^{z\,1,2,3}= 0.018, ~0.016, ~0.016$ at $J'/J = 0.1$).
In contrast, at intermediate regimes we find a quantum plateau
like magnetization  for $0.18 < J'/J \lesssim 1$
({\em c.f.}\ $S^{z\,1,2,3}=0.470, ~0.015, ~0.015$ at $J'/J = 0.5$), as well as an up-up-down
magnetization for modulated regions $J'/J \gtrsim 1$
({\em c.f.}\ $S^{z\,1,2,3}=-0.075, ~0.288 , ~0.288$ at $J'/J = 1.2$).
We find clear signals of a level crossing at
$J'/J = 1$: the ground state magnetization profiles correspond to
a quantum plateau ($S^z \approx 0,1/2,0$ at each trimer) for $0.18 < J'/J \lesssim 1$,
and up-up-down states (sign($S^z $) = -,+,- at each trimer) for $J'/J \gtrsim 1$.
One can also appreciate a sudden magnetization change at the level crossing point $J'/J = 0.18$;
the ground state magnetization profile corresponds to
uniform magnetization ($S^z \approx 0.166$ at each site) for $0 < J'/J < 0.18$ and to the quantum plateau
for $0.18< J'/J $.
For the 18 spins system the magnetization pattern is similar, except for the smooth behavior between
the uniform magnetization and quantum plateau phases; this corresponds to the
above mentioned possibility of the transition
being BKT-like instead of first order.

The numerical results can be summarized by an order parameter
\beq
M_s=\sum_i\cos(\frac{2\pi}{3}(i-2))<S^z_i>
\label{orderparameter}
\eeq
describing a modulated local magnetization with period three. As shown in Fig.\ \ref{profile} (lower panel),
this parameter allows the identification of three different phases.
Notice that the second order transitions, characterized by long range fluctuations,
are not expected to show up clearly in  finite size systems.
Thus the results around $J'/J \lesssim 1$ are compatible with our previous semiclassical analysis.

As a separate issue, we have also tested the confidence of the first order block-spin perturbation results.
To this end, we computed the spectrum of the Hamiltonian obtained in Eq.\ (\ref{Heff1}),
adapted to a finite size system of 24 spins with periodic boundary conditions, through the Jordan-Wigner
mapping to spinless fermions
(care has to be taken in imposing periodic (antiperiodic) boundary conditions
for odd (even) fermion filling to the resulting tight-binding Hamiltonian).
The  spectrum thus obtained can be compared with exact Lanczos diagonalization of the full spin system
at different values $J'/J$. As it is known, the convergence of this method to highly excited states
becomes progressively slow. To avoid this problem we restarted the
Lanczos procedure with an initial random state chosen orthogonal to
each of the eigenlevels previously found.
\begin{figure}[htbp]
    \centering
        \rput(2.8cm,4.5cm){(a)}
        \rput(2.6cm,-0.98cm){(b)}
        \includegraphics[width=0.5\textwidth]{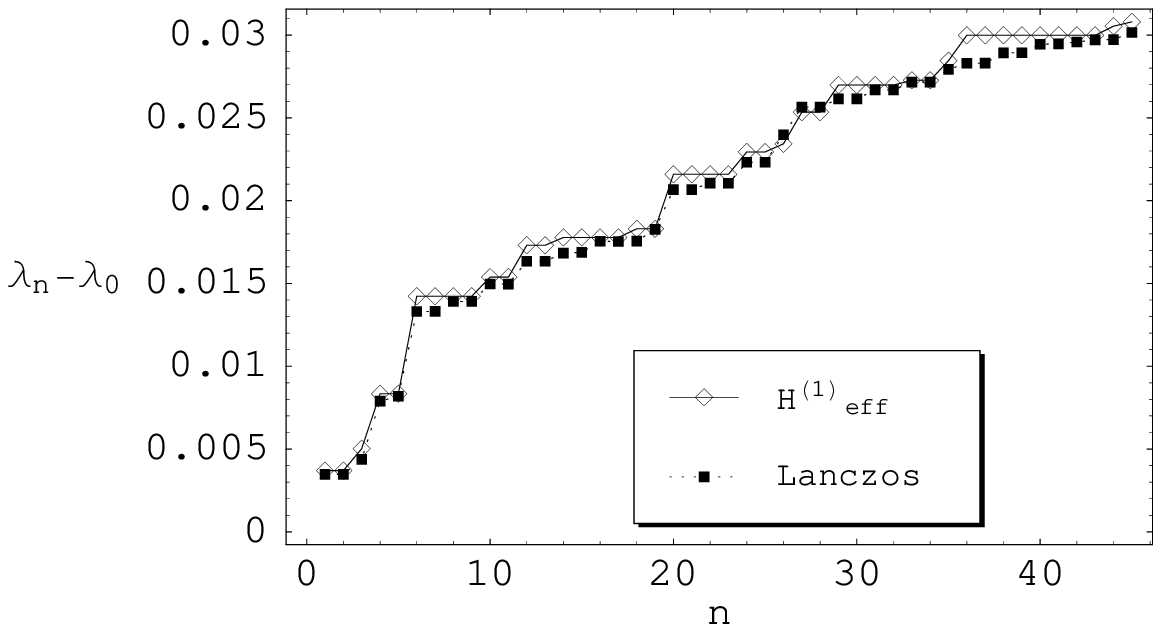}
        \includegraphics[width=0.5\textwidth]{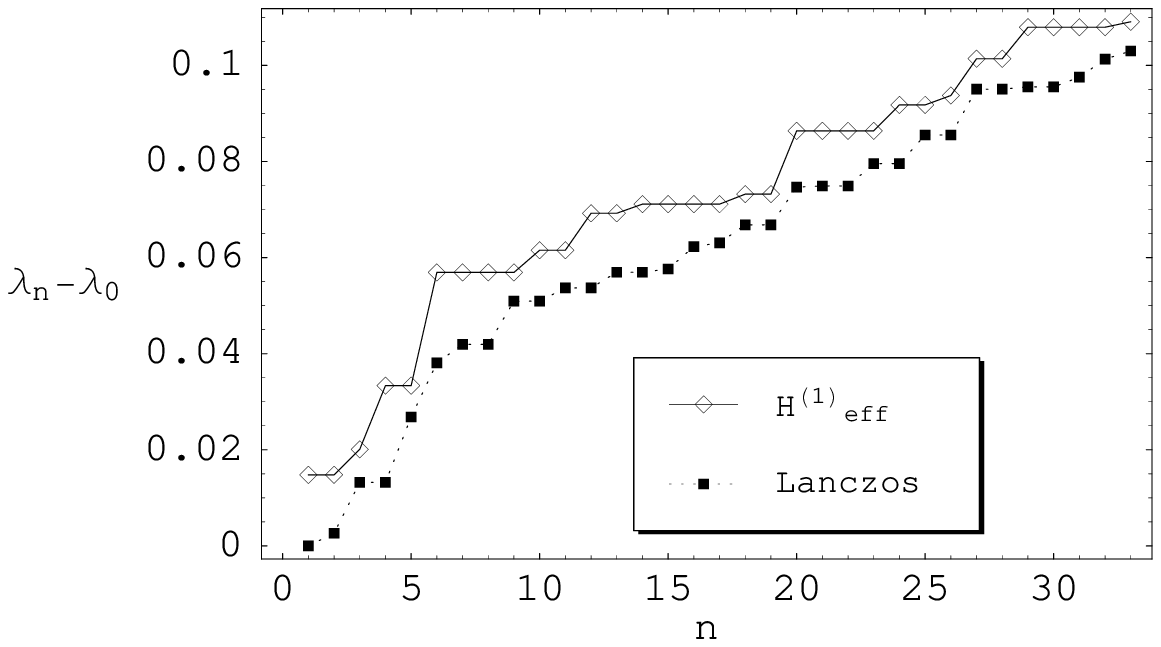}
    \caption{Comparison of non magnetic excitations $\lambda_n$ above the $M=1/3$ plateau,
    obtained on a finite system from the first order
    block-spin perturbation and from exact diagonalization:
    (a) $J'/J=0.05$,
    (b) $J'/J=0.2$.
    The horizontal and vertical axes are the excitation level (counting degeneracy) and
    $\lambda_n-\lambda_0$ respectively.
    }
    \label{exci005}
\end{figure}
In Fig.\ \ref{exci005} we plot in the upper panel the
first excitations for $J'/J=0.05$, showing very good agreement both in values and degeneracy of the energy
levels.
In the lower panel the same is plotted for $J'/J=0.2$, showing important deviations of the
$XY$ picture from the exact results. This deviation is expected because of the level crossing
found numerically at $J'/J=0.18$ in the 24 spins system.

\section{Summary and conclusions}

In the present note we have analyzed quantum phase transitions
in zig-zag antiferromagnetic spin $S=\frac12$ ladders at $M=1/3$ driven by
a trimerized modulation, as could be produced by adiabatic lattice deformations in a spin-Peierls like
transition.

Far from being stable, the triple
degeneracy of the $1/3$ magnetization plateau in homogeneous ladders is
lifted according to a triple sine-Gordon mechanism, giving place to several magnetic phases separated by
first and second order transitions. Numerical diagonalization results in finite systems
support our conclusions. A further transition
separates the weak modulation regime from the strong frustration regime.
From weak trimer coupling to beyond the homogeneous zig-zag point, the phase diagram is
schematically shown in Fig.\ \ref{schematicPhDiag}:
a phase with uniform magnetization described by non magnetic gapless chiral degrees of freedom,
then a transition to
a non degenerate quantum plateau state, then a second order transition,
in the Ising universality class, to a two-fold degenerate
state and finally a first order transition at the homogeneous point to a non degenerate
up-up-down ground state. The nature of the transition between the chiral gapless phase
and the quantum plateau is not resolved by our numerical data; we leave a blank in the phase diagram as
an open question on this issue.

Motivated by the spin-Peierls transition usually studied at zero magnetization,
it is interesting to investigate whether a lattice deformation
like the one analyzed in this work could take place at low temperatures in magnetized systems
because of the competition between magnetic energy gain and elastic energy loss.
We hope that the knowledge of the ground state structure and non magnetic excitations
presented here will be useful in such investigation. Related results will be published elsewhere.

{\em Acknowledgments:}
This work was partially supported by CONICET (Argentina),
ECOS-Sud Argentina-France
collaboration (Grant A04E03), PICS CNRS-CONICET (Grant 18294),
PICT ANCYPT (Grant 20350), and PIP CONICET (Grant 5037).
T.V.\ also acknowledges GNSF grant N 06-81-4-100.


\end{document}